\RequirePackage{fix-cm}
\documentclass[smallextended]{svjour3}       
\smartqed  
\usepackage{cite}
\usepackage{amsmath, amssymb, amsfonts}
\usepackage{algorithmic}
\usepackage{graphicx}
\usepackage{balance}
\usepackage{xcolor}
\usepackage{enumitem}
\usepackage{booktabs}
\usepackage{hyperref}
\definecolor{myPurple}{RGB}{120, 50, 120}
\hypersetup{
    colorlinks=true, 
    linkcolor=black, 
    citecolor=black, 
    urlcolor=myPurple
}

\usepackage{graphicx}
\usepackage{xspace}
\usepackage{amsmath, amssymb, amsfonts}
\usepackage{algorithmic}
\usepackage{graphicx}
\usepackage{textcomp}
\usepackage{makecell}
\usepackage{natbib}
\usepackage{caption}
\usepackage{listings}
\usepackage{seqsplit}

\begin{document}

\title{Understanding Typing-Related Bugs in Solidity Compiler
}


\author{Lantian Li  \and Yue Pan \and Dan Wang \and Jingwen Wu  \and Zhongxing Yu
}



\institute{Lantian Li \at
              72 Binhai Road, Jimo, Qingdao, P.R. China \\
              Shandong University \\
              \email{lilantian@mail.sdu.edu.cn}           
           \and
            Yue Pan \at
              72 Binhai Road, Jimo, Qingdao, P.R. China \\
              Shandong University \\
              \email{pany@mail.sdu.edu.cn}           
           \and
           Dan Wang  \at
              72 Binhai Road, Jimo, Qingdao, P.R. China \\
              Shandong University \\
              \email{wang.dan@mail.sdu.edu.cn} 
              \and
           Jingwen Wu \at
              72 Binhai Road, Jimo, Qingdao, P.R. China \\
             Shandong University \\
              \email{elowen.jjw@gmail.com}           
           \and
           Zhongxing Yu \at
              72 Binhai Road, Jimo, Qingdao, P.R. China \\
            Shandong University \\
              \email{zhongxing.yu@sdu.edu.cn}           
}

\date{Received: date / Accepted: date}

\maketitle

\begin{abstract}
The correctness of the Solidity compiler is crucial for ensuring the security of smart contracts. However, the implementation complexity of its type system often introduces elusive defects. This paper presents the first systematic empirical study on typing-related bugs in the Solidity compiler. To systematically analyze these bugs, we collected 146 officially confirmed and fixed typing-related bugs from the Solidity compiler’s official GitHub repository. For each bug, we conducted an in-depth analysis and classification from four dimensions: symptoms, root causes, exposure conditions, and fix strategies. Through this study, we reveal unique distribution patterns and key characteristics of such bugs, and summarize 12 core findings. We additionally give the implications of our findings, and these implications not only deepen the understanding of inherent weaknesses in the Solidity compiler but also provide new insights for detecting and fixing typing-related bugs in the Solidity compiler. 

\keywords{Solidity \and Typing-related bugs \and Empirical study}
\end{abstract}

\section{Introduction}
\label{intro}
Blockchain technology has given rise to a new generation of decentralized platforms, epitomized by Ethereum, whose core innovation lies in the introduction of smart contracts—programmed entities capable of automatically executing agreement terms without reliance on trusted third parties (\citealt{antonopoulos2018mastering}). Today, smart contracts have become the cornerstone of critical applications such as decentralized finance (\citealt{zhou2023sok}), supply chain traceability (\citealt{hamledari2021measuring}), and digital asset management (\citealt{hong2020fabasset}). They hold and manage vast amounts of digital assets (\citealt{DefiLlama, code4rena,inconsistentstateupdate}), making their security and reliability of paramount importance to the stability and trust of the entire blockchain ecosystem.

As the dominant programming language in the Ethereum ecosystem, Solidity serves as the foundation for the vast majority of smart contracts. It is a high-level, statically-typed language specifically designed for smart contracts, and the Solidity compiler is the core tool responsible for translating Solidity source code into bytecode executable on the Ethereum Virtual Machine (EVM). The correctness of the compilation process directly determines whether the behavior of the deployed on-chain contract aligns with developer expectations. If the compiler itself contains flaws, even logically correct source code may be compiled into bytecode with security vulnerabilities or functional errors. Historically, this has led to significant financial losses (\citealt{ma2024towards, mitropoulos2023syntax}), underscoring the critical importance of compiler quality for smart contract security.

In the field of software engineering, significant progress has been made in reliability research for foundational software such as compilers (\citealt{ren2021unleashing, gu2023llm, liu2023nnsmith}). Empirical studies based on systematic categorization and analysis of historical defects represent a key approach to understanding and improving compiler quality. However, existing research has primarily focused on traditional compilers like GCC and LLVM (\citealt{he2025bug, xu2023silent}), with limited attention given to domain-specific language compilers such as the Solidity compiler. The Solidity compiler possesses unique complexities due to its distinctive objectives (\emph{e.g.}, gas optimization, integration of formal verification) and runtime environment (blockchain virtual machines) (\citealt{buterin2013ethereum}). In particular, the type system of Solidity is expressive (with dependence on complex type theories or typing features such as parametric polymorphism and type inference) and has sophisticated interactions with other compilation phases such as front-end analysis and optimization processes, thus the implementation of type systems is extremely intricate and can harbor complex and unique defect patterns. 
While recent empirical work has broadly characterized defects in the Solidity compiler from a macroscopic perspective (\citealt{ma2024towards}), a dedicated in-depth investigation into typing-related defects remains absent.

To address this research gap, this paper conducts the first systematic empirical study on typing-related bugs in the Solidity compiler, providing an in-depth analysis of their symptoms, root causes, exposure conditions, and fix strategies. Specifically, we aim to explore the following 4 research questions:

\begin{itemize}[leftmargin=*]
\item {\textbf{RQ1 (Symptoms): What are the main symptoms of typing-related compiler bugs?} This question aims to elucidate the impact of typing-related bugs.}
\item {\textbf{RQ2 (Root Causes): What are the root causes of typing-related compiler bugs?} This question seeks to identify the underlying factors leading to typing-related compiler bugs.}
\item {\textbf{RQ3 (Exposure Conditions): What conditions are required to expose typing-related compiler bugs?} This question aims to study how typing-related compiler bugs can be triggered.} 
\item {\textbf{RQ4 (Fix Strategies): How do developers fix such typing-related compiler bugs to prevent harm?} 
This question aims to explore effective fix strategies.}
\end{itemize}

To investigate these 4 questions, we examine 146 typing-related bugs in the Solidity compiler documented before April 8, 2025. These bug samples are collected from the Solidity compiler's official GitHub repository, all of which have been confirmed and fixed. For each of the 146 bugs, we conduct an in-depth analysis of the contract code triggering the bug, error reports, fixes, and code comments to thoroughly understand their symptoms, root causes, exposure conditions, and effective fix strategies. Our large-scale study enables us to deliver 12 original and important findings. We also give the implications of our findings, which provide insights for developers, researchers, tool builders, and language or library designers to improve various aspects of the Solidity compiler.

The main contributions of this paper are as follows:
\begin{itemize}[leftmargin=*]
\item {We conduct the first systematic empirical study on typing-related bugs in the Solidity compiler.}
\item { By investigating 146 typing-related bugs, we provide an in-depth analysis on diverse aspects, including bug symptoms, root causes, exposure conditions, and fix strategies. }
\item {We derive 12 original and significant findings with actionable implications, which benefit developers, researchers, tool builders, as well as language or library designers involved in Solidity compiler engineering.
}
\end{itemize}

The remainder of this paper is structured as follows. We first present necessary background in Sec~\ref{Background}. Sec~\ref{Methodology} explains the methodology to perform the empirical study. Sec~\ref{Symptoms}, Sec~\ref{rootcause}, Sec~\ref{Exposure-Conditions}, and Sec~\ref{fixstrategy} present the results of the empirical study about the symptoms, root causes, exposure conditions, and fix strategies for typing-related compiler bugs respectively. The last two sections give some closely related work and the conclusion respectively.

Our replication package (including code, dataset, etc.) is available at \url{https://github.com/SolComTyp/SolTypeStudy}

\section{Background}
\label{Background}
In this section, we present essential background on smart contracts, the Solidity compiler, and typing-related bugs.

\subsection{Smart Contract}
A smart contract is a program operating on a distributed ledger that automatically executes corresponding contractual terms when predefined conditions are met (\citealt{antonopoulos2018mastering}). The development of smart contracts typically employs specially designed high-level programming languages, among which Solidity is the most prominent due to its rich feature set and mature toolchain (\citealt{Soliditylang,liIDOL,Solsmith}). Solidity adopts a JavaScript-like syntax and supports complex contract-oriented programming paradigms, including inheritance, library integration, and user-defined types. Each smart contract comprises a set of state variables and functions that collectively define its behavioral logic
and data storage mechanisms.
During deployment, the contract source code is compiled into specific EVM bytecode, which is then reliably executed across decentralized networks through distributed consensus mechanisms. 
This unique technical architecture enables smart contracts to show significant potential in domains requiring trusted collaboration mechanisms, such as digital finance and decentralized governance (\citealt{zou2019smart}). 

\subsection{Solidity Compiler}
The Solidity compiler, as a core component of the smart contract development toolchain, undertakes the critical task of transforming high-level Solidity code into executable EVM bytecode (\citealt{liu2023empirical}). This compiler adopts a multi-layer processing architecture: the front-end performs lexical analysis, syntax parsing, and semantic analysis to construct a complete Abstract Syntax Tree (AST); the middle-end (through optimization passes based on the Yul intermediate language) (\citealt{Yulsolidity}) restructures and optimizes contract logic with a focus on reducing gas consumption (\citealt{vacca2022empirical}) and bytecode size; finally, the back-end generates EVM-compliant bytecode. Beyond basic compilation, the compiler also produces the Application Binary Interface (ABI)—a standardized interface description that enables external applications to correctly identify and invoke contract functions with their parameter types (\citealt{tong2025sbugchecker}). Notably, the Solidity compiler integrates the formal verification tool SMTChecker, which performs static verification of mathematical properties during compilation and provides additional assurance for safety-critical code (\citealt{mitropoulos2024broken}). These unique functional modules, combined with storage models and exception handling mechanisms specifically designed for blockchain environments, distinguish the Solidity compiler in terms of architectural complexity and functional specificity compared to traditional compilers (\citealt{alt2022solcmc}), thus introducing new research challenges.

\subsection{Typing-Related Bugs}
\label{TypeRelatedBugs}
\emph{Typing-related bugs refer to implementation defects that arise during the compiler's execution of type information processing operations} (\citealt{chaliasos2022finding, sotiropoulos2024api}), and constitute an important bug category in compiler research (\citealt{chaliasos2021well}). Typing-related bugs can span various compilation phases, and prevent the compiler from properly performing core operations such as type representation, conversion, inference, verification, and propagation.
For the Solidity compiler, typing-related bugs exhibit unique complexities. In particular, the implementation of type system rules requires coordinated handling of blockchain-specific storage models (\citealt{li2018blockchain}), gas optimization mechanisms, and deep integration with formal verification components. 
Thus, systematic categorization and investigation of these typing-related bugs not only facilitates a deeper understanding of the Solidity compiler's internal mechanisms, but also provides crucial support for enhancing the security and reliability of smart contracts.

\section{Methodology}
\label{Methodology}
\subsection{Subjects and Data Used in Our study}
The data for this study is sourced from the official GitHub repository of the Solidity compiler. We now detail the data collection process. First, it is the standard practice for Solidity compiler developers to include labels in issue descriptions. We thus retrieved all closed issues labeled as ``bug" up to April 8, 2025 through the GitHub API, obtaining 1386 initial samples. We then examined whether these issues referenced related pull requests, thereby filtering out bugs without explicit fix commits and yielding a total of 614 remaining bugs. Finally, we conducted a detailed manual examination for each fixed bug. By closely examining developer descriptions, issue discussion records, 
and fix solutions, we strictly evaluated them against the criteria of typing-related bugs given in Sec.~\ref{TypeRelatedBugs}, keeping bugs related to type system implementation. As a result, we identified 146 qualified typing-related compiler bugs as the subjects for this study. These bug samples form a high-quality dataset, with each containing bug descriptions, code or operations triggering the bug, and fix solutions. 

\subsection{Study Methodology}
For each typing-related compiler bug, we thoroughly analyze the error reports, the Solidity source code triggering the bug, the compiler's fix code and its corresponding commit message, as well as the code comments, to fully understand the bug's symptoms, root causes, exposure conditions, and fix strategies. To minimize subjective bias, two authors of this paper independently analyzed each bug. Subsequently, we categorized and described the typing-related compiler bugs along four dimensions: symptoms, root causes, exposure conditions, and fix strategies. It is worth mentioning that our researchers possess at least three years of experience in Solidity smart contract development and compiler testing research.


The manual inspection is in particular composed of two phases. In the first phase, two independent authors of the paper derive an initial result after the analysis. 
After the classification, the Cohen’s kappa coefficient is calculated (\citealt{cohen1960coefficient}), which is a statistic for measuring inter-rater reliability for qualitative items. The Cohen's kappa coefficients for symptoms, root causes, exposure conditions, and fix strategies are 0.88, 0.78, 0.85, and 0.86 respectively, indicating good agreement rates. In the second phase, the two authors compare the initial results derived, have a discussion about the disagreements, and refine their initial results according to the outcome of the discussion (\emph{i.e.}, the consensus that has been reached). Note that Cohen’s kappa coefficients and multi-phase manual inspections have been widely used by existing works (\citealt{li2023understanding, ICSE43902202100019, achamyeleh2025bridging, yuannotation}).

\subsection{Threats to Validity}
\textbf{External Validity.} The major threat to external validity lies in the representativeness of the selected bugs. To mitigate this threat, our study focuses on bugs accompanied by fixes. These fixed bugs represent genuine bugs and are important to developers (as they have been fixed).
Besides, during the selection process, we specifically chose issues explicitly labeled with the ``bug" tag. We avoided using keywords in our search to reduce the chance of missing relevant bugs.
Future works will use more bugs to further reduce this threat. 

\textbf{Internal Validity.} The internal validity is primarily related with the manual processes that may introduce bias. To mitigate this threat, we followed a well-established and widely-used procedure where each manual analysis was performed independently by two experienced authors of this paper. Their independent findings were subsequently compared. Specifically, for the manual inspection of the 146 typing-related bugs, they examined error reports, contract code triggering the bugs, the compiler's fix code and its commit messages, along with code comments to gain deep insights into the bugs. Furthermore, the complete artifact of this paper is available online, providing readers with deeper insights into our research and analysis.

\section{Symptoms for Typing-related Bugs (RQ1)}
\label{Symptoms}

This section presents our research results on the symptoms of typing-related bugs in the Solidity compiler. As shown in Table~\ref{Symptoms}, the symptoms observed for the 146 typing-related bugs can be classified into six categories, .

\begin{table}
\caption{Category of the symptoms for typing-related bugs.}
\begin{center}
\begin{tabular}{ l c c }
\toprule
\textbf{Symptoms categories} & \textbf{Instance} & \textbf{Percentage}  \\
\midrule
Abort Without Diagnostic Message & 60 & 41.10\% \\
False Rejection & 58 & 39.73\% \\
Missed Errors & 12 & 8.22\% \\
Incorrect Output & 9 & 6.16\% \\
Misleading Diagnostics & 5 & 3.42\% \\
Hanging & 2 & 1.37\% \\
\bottomrule
\end{tabular}
\label{Symptoms}
\end{center}
\end{table}

\subsection{Abort Without Diagnostic Message}
The symptom of abort without diagnostics message manifests when the Solidity compiler encounters certain code containing type-related issues, the compilation process aborts and normal diagnostic information is not provided. 
The compiler instead can only output obscure messages or no prompt at all, and may also fail to generate target code or produce failures such as internal compiler errors. This category of errors generally originates from inadequate handling of complex or illegal code patterns, including recursive type definitions, out-of-range literals, invalid tuple operations, intricate inheritance relationships, or unsupported type combinations. The compiler's internal defense mechanisms are insufficient to properly handle these exceptional cases, leading to failures in type system inference, memory management errors, or infinite recursion (\emph{e.g.}, \href{https://github.com/argotorg/solidity/issues/127}{ID: 127}, \href{https://github.com/argotorg/solidity/issues/561}{561}). This symptom not only obscures the nature of the original code issues but also places developers in a challenging debugging position, thus revealing significant deficiencies in the compiler's robustness and error recovery mechanisms.

\begin{center}
\fcolorbox{black}{gray!25}{\parbox{0.97\linewidth}{
\textit{\noindent\textbf{Finding 1}: 
The error diagnosis mechanisms lack contextual awareness of the type system, and the compiler frequently aborts without diagnostic message when compiling Solidity source code with type-related issues (41.10\%).
}

\textit{
\noindent\textbf{Implication 1}: For testing techniques like fuzzers, it is important to guide them towards generating Solidity  code with type-related issues in order for detecting typing-related compiler bugs.
For IDE developers, it calls for developing intelligent type diagnosis assistance tools that offer supplementary explanations based on known defect patterns.
}}}
\end{center}

\subsection{False Rejection}
The false rejection symptom manifests when the Solidity compiler incorrectly rejects well-typed and semantically valid programs, generating misleading error messages or internal failures. This behavior typically stems from issues such as an overly restrictive type system, failures in overload resolution, internal state inconsistencies, or false positives in unimplemented features. Specific instances include the compiler's inability to correctly resolve function overload signatures, improper handling of constant dependencies, misjudgment of data location compatibility, 
or assertion failures caused by type inference errors in static analyzers like the SMTChecker (\emph{e.g.}, \href{https://github.com/argotorg/solidity/issues/621}{ID: 621}, \href{https://github.com/argotorg/solidity/issues/1709}{1709}). These scenarios prevent otherwise compilable programs from being processed, accompanied by spurious messages such as ``type mismatch" and ``internal compiler error".
This symptom not only obscures the actual correctness of the program but also increases debugging complexity, revealing logical flaws in the compiler's semantic analysis, type checking, or model verification phases.

\subsection{Missed Errors}
The symptom of missed errors manifests when the Solidity compiler silently accepts invalid or erroneous programs that should be rejected, failing to generate necessary error or warning messages. This allows defective code to pass compilation checks while potentially causing undefined behavior, security vulnerabilities, or language specification violations during runtime. Such defects typically originate from incomplete type system checks, missing semantic rule validations, or blind spots in static analysis. Representative cases include: permitting illegal exponentiation operations on signed integers, accepting negative-value address conversions, or ignoring the illegal presence of empty types in tuples
(\emph{e.g.}, \href{https://github.com/argotorg/solidity/issues/1343}{ID: 1343}, \href{https://github.com/argotorg/solidity/issues/9857}{9857}). These omissions enable serious errors—which should have been caught during compilation, such as violations of EVM opcode specifications 
and storage pointer misuses—to remain latent until runtime. This not only compromises type safety and contract reliability but also creates exploitation opportunities that can lead to asset losses, 
thereby revealing significant deficiencies in the compiler's enforcement of semantic constraints and the coverage of its static verification.

\begin{center}
\fcolorbox{black}{gray!25}{\parbox{0.97\linewidth}{
\textit{\noindent\textbf{Finding 2}: 
The error messages for typing-related bugs demonstrate insufficient quality, with numerous instances of false rejection and missed errors (47.95\%). 
}

\textit{
\noindent\textbf{Implication 2}: 
For language designers, it is beneficial to simplify typing rules, clarify edge cases, and avoid designs that easily cause confusion.
}}}
\end{center}

\subsection{Incorrect Output}
For incorrect output symptom, the Solidity compiler receives valid source programs and successfully compiles them, but the generated output deviates from expectations. The output includes, but is not limited to, SMT-LIB scripts, bytecode, return values, ABIs, and ASTs. These errors may lead to runtime data persistence issues, type confusion, behavioral inconsistencies, or verification failures. For example, the compiler may incorrectly assign storage variables to memory locations, 
ignore event overloading during ABI generation, 
or generate invalid SMT-LIB scripts in static analysis tools like SMTChecker (\emph{e.g.}, \href{https://github.com/argotorg/solidity/issues/11631}{ID: 11631}, \href{https://github.com/argotorg/solidity/issues/14375}{14375}). 
Ultimately, these defects cause the generated executables or metadata to exhibit erroneous behavior during runtime rather than producing the expected correct results.

\subsection{Misleading Diagnostics}
The symptom of misleading diagnostics manifests when the Solidity compiler successfully detects anomalies in programs and generates diagnostic information, but the provided error messages or warnings are inaccurate, incomplete, or misleading. This prevents developers from quickly identifying the actual issues and may even lead to incorrect fixes. Such errors typically originate from biases in the compiler's internal type inference algorithms, flaws in error message generation logic, or insufficient context awareness. For instance, the compiler may erroneously suggest that data supporting only the memory location can arbitrarily choose between memory or storage (\emph{e.g.}, \href{https://github.com/argotorg/solidity/issues/11405}{ID: 11405}). These misleading diagnostics not only fail to assist in problem resolution but may also obscure genuine semantic errors, forcing developers into debugging difficulties and potentially introducing new defects due to reliance on incorrect suggestions.

\subsection{Hanging}
Among typing-related bugs, the hanging symptom occurs infrequently. Hanging manifests when the Solidity compiler enters an infinite loop or deadlock state while processing certain code patterns, causing the compilation process to become unresponsive and unable to terminate normally. This category of errors is typically triggered by extreme code patterns, such as 
when handling recursive definitions like constant self-references, the compiler may fail to complete the evaluation process and enter logical closed loops, causing the entire compilation task to hang (\emph{e.g.}, \href{https://github.com/argotorg/solidity/issues/10732}{ID: 10732}). This not only consumes substantial computational resources but also forces developers to terminate the process through external means. 

\section{Root Causes for Typing-related Bugs (RQ2)}
\label{rootcause}
This section presents our research results on the root causes of typing-related bugs in the Solidity compiler. As shown in Table~\ref{Causes}, the root causes of the 146 typing-related bugs can be classified into five categories. 

\begin{table}
\caption{Category of the root causes for typing-related bugs.}
\begin{center}
\begin{tabular}{ l c c }
\toprule
\textbf{Root causes categories} & \textbf{Instance} & \textbf{Percentage}  \\
\midrule
Core Type Operation Defects & 57 & 39.04\% \\
Semantic Analysis Defects & 51 & 34.93\% \\
Symbol and Scope Management Failures & 18 & 12.33\% \\
AST and IR Transformation Errors & 13 & 8.91\% \\
Defects in Error Handling Mechanisms & 7 & 4.79\% \\
\bottomrule
\end{tabular}
\label{Causes}
\end{center}
\end{table}

\subsection{Core Type Operation Defects}
Core type operation defects are directly related with the implementation of basic typing mechanisms such as type conversion, comparison, and inference, and contribute to the largest proportion of typing-related compiler bugs (with the percentage 39.04\%). These bugs fundamentally arise when the compiler fails to strictly adhere to formal rules or adequately handle the language's complex features during basic type system operations. They are further divided into three subcategories as follows.

\textbf{a. Incorrect Type Conversion.}
This subtype accounts for 18.49\% of the total 146 typing-related bugs. Type conversion errors in the Solidity compiler reveal deficiencies in its type system when handling implicit and explicit type conversions, particularly evident in scenarios involving complex type hierarchies, ABI encoding rules, and data location semantics. These bugs primarily manifest as missing conversion rules, inconsistent data location handling, and inadequate support for custom types.
When processing the conversion of literals to specific types, such as converting string literals to \texttt{address} types, the compiler misidentifies string literals as storage types, thereby allowing illegal conversions (\emph{e.g.}, \href{https://github.com/argotorg/solidity/issues/4535}{ID: 4535}). Moreover, in ABI encoding contexts, the compiler's support for converting fixed-point number types, enumeration types, and function parameter types remains incomplete. It fails to correctly handle implicit conversion rules for \texttt{ufixed}/\texttt{fixed} types or the underlying \texttt{uint} representation of enums, triggering internal errors during \texttt{abi.encode} operations or event generation
(\emph{e.g.}, \href{https://github.com/argotorg/solidity/issues/4715}{ID: 4715}).
Data location consistency (\emph{e.g.}, compatibility between \texttt{calldata} and \texttt{memory}) presents another critical issue. During function overriding and callback function passing, 
the compiler neglects strict validation of data location matching, resulting in failed implicit conversions 
(\emph{e.g.}, \href{https://github.com/argotorg/solidity/issues/12718}{ID: 12718}).
Additionally, for custom types and complex type operations (such as type conversions of mapping keys or indirect calls to storage arrays' pop method), the type conversion mechanism lacks sufficient contextual awareness. This prevents correct invocation of wrapper functions or proper handling of type wrapping/unwrapping operations (\emph{e.g.}, \href{https://github.com/argotorg/solidity/issues/15888}{ID: 15888}, \href{https://github.com/argotorg/solidity/issues/10870}{10870}).
The introduction of the SMT checker further exposes integration flaws between type conversion and model verification. When processing conversions from string literals to byte types or mapping comparisons, 
the SMT checker triggers logic exceptions due to inconsistent type sorting 
or erroneous expression generation 
(\emph{e.g.}, \href{https://github.com/argotorg/solidity/issues/15813}{ID: 15813}, \href{https://github.com/argotorg/solidity/issues/1709}{1709}).

\begin{figure}[htbp]
\centering
\includegraphics[width=0.53\textwidth]{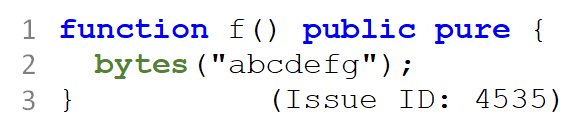}
\caption{An example that triggers an incorrect type conversion bug.}
\label{fig:conversation}
\end{figure}

Fig.~\ref{fig:conversation} shows a code example that triggers a type conversion bug. When explicitly converting the string literal ``abcdefg" to the bytes type, the compiler mistakenly interprets it as a storage pointer instead of the expected memory type. This leads to the misclassification of a legally permissible type conversion as an illegal operation. 

\begin{center}
\fcolorbox{black}{gray!25}{\parbox{0.97\linewidth}{
\textit{\noindent\textbf{Finding 3}: 
18.49\% of the typing-related bugs arise due to type conversion errors in the Solidity compiler, particularly evident in scenarios involving complex type hierarchies, ABI encoding rules, and data location semantics.
}

\textit{
\noindent\textbf{Implication 3}: Compiler developers need to establish specialized ABI type compatibility test suites, and contract developers should avoid using these error-prone types in critical business logic.
Besides, IDE developers can create data location flow analysis tools to provide early warnings during the coding phase.
}}}
\end{center}

\textbf{b. Incorrect Type Comparison and Type-related Constraint Calculation.}
This subtype accounts for 12.33\% of the total 146 typing-related bugs. Type comparison and constraint calculation errors in the Solidity compiler reveal inherent weaknesses in its type system when handling complex type relationships, numerical boundaries, and storage layouts. The essence of such bugs lies in the compiler's failure to adequately adhere to the formal rules of the type system during type compatibility checks (\emph{e.g.}, subtype relationship validation in assignments) and type constraint calculations (\emph{e.g.}, storage offset calculations).
At the type comparison level, the compiler often lacks sufficient contextual awareness or termination conditions when comparing special keywords (\emph{e.g.}, \texttt{super}) with regular contract types, judging compatibility of multi-dimensional array types, 
or performing equivalence checks on recursive type structures. 
This often leads to infinite recursion or violations of internal assertions (\emph{e.g.}, \href{https://github.com/argotorg/solidity/issues/7448}{ID: 7448}).
In type-related constraint calculation, the compiler exhibits deficiencies in handling numerical extreme values (\emph{e.g.}, excessively large integers).
It frequently encounters numerical overflows or underflows when computing array sizes, exponentiation results, or storage layouts, while lacking effective exception handling 
strategies (\emph{e.g.}, \href{https://github.com/argotorg/solidity/issues/15899}{ID: 15899}). Meanwhile, the compiler shows obvious shortcomings in calculating memory layouts for complex types (\emph{e.g.}, ABI encoding of struct return values exceeding 32 bytes), failing to properly reconcile the abstract rules of the type system with the concrete limitations of the EVM (such as the 32-byte load restriction) (\emph{e.g.}, \href{https://github.com/argotorg/solidity/issues/3069}{ID: 3069}).

\begin{figure}[htbp]
\centering
\includegraphics[width=0.82\textwidth]{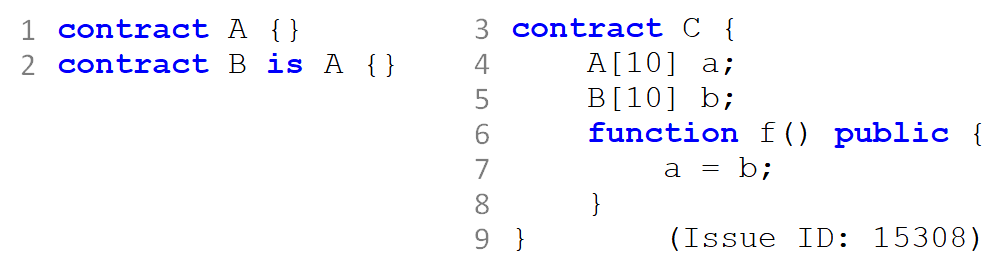}
\caption{An example that triggers an incorrect type comparison bug.}
\label{fig:Comparison2}
\end{figure}

Fig.~\ref{fig:Comparison2} gives a code example that reveals a flaw in the Solidity compiler's type comparison logic. When attempting to assign a subtype array \texttt{B[10]} to a parent type array \texttt{A[10]}, the SMT model checker within the compiler incorrectly determines that the two are incompatible, consequently reporting an SMT logic error. This exposes incorrect handling in the compiler's internal type system implementation regarding type compatibility checks for arrays within inheritance hierarchies.

\begin{center}
\fcolorbox{black}{gray!25}{\parbox{0.97\linewidth}{
\textit{\noindent\textbf{Finding 4}: 
12.33\% of the typing-related bugs arise due to type comparison and type-related constraint calculation errors in the Solidity compiler when handling complex type relationships, numerical boundaries, and storage layouts.
}

\textit{
\noindent\textbf{Implication 4}: Verification tools need to enhance detection capabilities for type relation violation, arithmetic overflow/underflow, and storage layout violation. Besides, compiler developers should introduce safe mathematical libraries to handle all boundary calculations.
}}}
\end{center}

\textbf{c. Incorrect Type Inference.}
This subtype accounts for 8.22\% of the total 146 typing-related bugs. This type of bug arises from incorrect type inference within the Solidity compiler. Specifically, for a given typed variable, the compiler either deduces an incorrect type or fails to infer any type at all. The root cause lies in systematic deficiencies within the type system when handling the language's complex features, primarily manifesting as incomplete construction of type constraints, failure in type information propagation, and inadequate recursive processing mechanisms. These deficiencies cause the type inference engine to fail in correctly initializing type annotations or constructing effective constraint systems when processing custom type wrappings (\emph{e.g.}, user-defined types based on elementary types), nested type structures (\emph{e.g.}, mappings containing struct keys), or dynamic type encodings (\emph{e.g.}, parameters involving internal types). Consequently, this leads to null pointer accesses, stack overflows, or type assertion failures. For instance, when tuple assignment elements mismatch or array length expressions remain uninitialized, the type checker fails to properly populate type information, causing subsequent stages (like the SMTChecker) to crash upon accessing uninitialized pointers (\emph{e.g.}, \href{https://github.com/argotorg/solidity/issues/5875}{ID: 5875}). 
Furthermore, the compiler's insufficient capability in type resolution for complex expressions (such as array length calculations within events) 
and operator overloading 
(custom operators implemented via \texttt{using} statements) 
further exposes the fragility of its type inference when integrated with semantic checks 
(like constant initialization validity)
and model verification (\emph{e.g.}, \href{https://github.com/argotorg/solidity/issues/9434}{ID: 9434}).

\begin{figure}[htbp]
\centering
\includegraphics[width=0.75\textwidth]{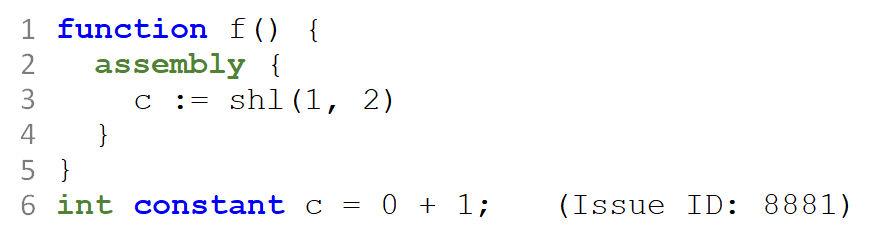}
\caption{An example that triggers an incorrect type inference bug.}
\label{fig:inference}
\end{figure}

Fig.~\ref{fig:inference} gives a code example that triggers an incorrect type inference bug. When an inline assembly statement attempts to reference or modify a constant \texttt{c} defined after the function \texttt{f()}, the compiler cannot obtain the type information of this constant during the type inference phase, consequently throwing an internal error of ``Type requested but not present".

\subsection{Semantic Analysis Defects}
Semantic analysis defects reveal weaknesses in the Solidity compiler's comprehension of code semantics, and account for 34.93\% of all typing-related bugs. The fundamental issue lies in the compiler's failure to establish a reliable mapping from syntactic structures to semantic constraints. Whether due to missing critical validation checks or design flaws in core analysis mechanisms, these deficiencies prevent the compiler from correctly identifying code patterns that violate language specifications or EVM constraints.
Such bugs may trigger cascade effects - when erroneous semantic interpretations propagate to subsequent compilation phases, they often cause more severe internal crashes or generate unsafe bytecode. They are further divided into two subcategories as follows.

\textbf{a. Missing Validation Checks.}
This subtype accounts for 23.29\% of the total 146 typing-related bugs. Missing validation checks reflect omissions in critical constraint verification during the semantic analysis phase. These issues cause the compiler to incorrectly accept code that violates language specifications or virtual machine constraints, subsequently leading to compilation anomalies, runtime errors, or security vulnerabilities. The essence of this problem lies in the compiler's failure to establish comprehensive verification barriers to intercept semantically invalid code structures.
At the type system level, the compiler lacks termination checks for recursive type definitions (\emph{e.g.}, self-referential structs), resulting in infinite recursion and stack overflow (\emph{e.g.}, \href{https://github.com/argotorg/solidity/issues/9443}{ID: 9443}). Regarding the storage model, it fails to validate the correctness of data locations (\emph{e.g.}, misusing mapping types for memory arrays, \href{https://github.com/argotorg/solidity/issues/5518}{ID: 5518}) or the initialization state of storage pointers (\emph{e.g.}, uninitialized \texttt{storage} pointers, \href{https://github.com/argotorg/solidity/issues/10821}{ID: 10821}), causing disorganized storage layouts.
For language-specific constructs, the compiler lacks boundary checks for scenarios such as constant self-references and compatibility with experimental features (\emph{e.g.}, ABIEncoderV2) (\citealt{abiencoderv2}), 
allowing unconventional usage patterns to bypass semantic constraints.
At the contract architecture level, missing validation of constructor parameter passing, applicability of the \texttt{override} modifier (\emph{e.g.}, state variables attempting to override functions), and consistency between function visibility and state modifications leads to broken inheritance hierarchies and interface contracts (\emph{e.g.}, \href{https://github.com/argotorg/solidity/issues/6901}{ID: 6901}).
More critically, these missing validations often trigger chain reactions rather than simple error reporting omissions. When invalid code proceeds to subsequent compilation stages (\emph{e.g.}, Yul code generation), 
violated preconditions trigger more severe internal assertion failures or logic errors. 
This fragmentation of verification mechanisms exposes the compiler's incompleteness in enforcing language specifications.

\begin{figure}[htbp]
\centering
\includegraphics[width=0.64\textwidth]{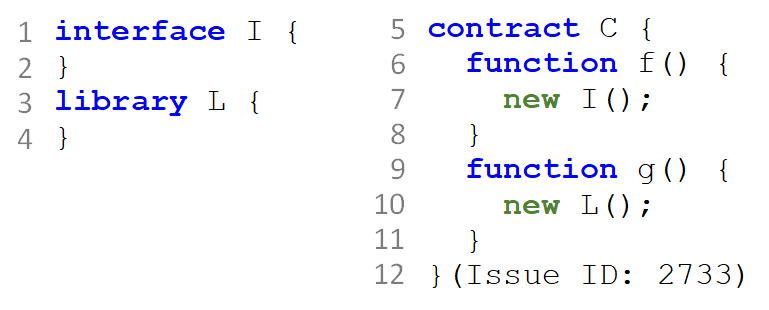}
\caption{An example that triggers a missing validation check bug.}
\label{fig:check2}
\end{figure}

Fig.~\ref{fig:check2} gives an  example code snippet that triggers a missing validation check bug. The example defines \texttt{interface I} and \texttt{library L}, but attempts to execute \texttt{new I()} and \texttt{new L()} in functions \texttt{f()} and \texttt{g()} of contract \texttt{C}, respectively. According to language specifications, neither interfaces nor regular libraries can be directly instantiated. The compiler should perform type checking during semantic analysis and report errors accordingly. However, this code is incorrectly accepted by the compiler, indicating the absence of necessary validation checks. 

\begin{center}
\fcolorbox{black}{gray!25}{\parbox{0.97\linewidth}{
\textit{\noindent\textbf{Finding 5}: 
Missing validation checks account for 23.29\% of all typing-related bugs in our study, indicating that the compiler lacks a unified semantic constraint network. 
}

\textit{
\noindent\textbf{Implication 5}: Compiler maintainers need to establish a cross-module collaborative verification framework. Besides, security auditors should pay special attention to patterns corresponding to these verification blind spots.
}}}
\end{center}

\textbf{b. Incorrect Analysis Mechanisms.}
This subtype accounts for 11.64\% of the total 146 typing-related bugs. Incorrect analysis mechanisms stem from design flaws in the core algorithms' parsing logic for language constructs, preventing the compiler from accurately establishing abstract representations of code structures. During the analysis process, the compiler misinterprets the categories, semantic roles, or contextual relationships of syntactic elements, causing subsequent analysis steps to build upon erroneous premises. For instance, the compiler may confuse the semantic hierarchy between destructuring syntax and assignment operations when handling nested tuple assignments (\emph{e.g.}, \href{https://github.com/argotorg/solidity/issues/8450}{ID: 8450}). Moreover, when dealing with interactions between inline assembly and advanced types, it may potentially permit access to restricted contract functions within assembly blocks (\emph{e.g.}, \href{https://github.com/argotorg/solidity/issues/8386}{ID: 8386}).
The fundamental issue with these incorrect analysis mechanisms lies in the inconsistency across the compiler's various analysis phases (lexical analysis, syntactic analysis, semantic analysis) in interpreting the same language constructs, coupled with inadequate intermediate representations to fully capture the mapping relationship from syntax to semantics. This ultimately causes the analysis pipeline to lose consistency when encountering edge-case syntactic structures or cross-feature interactions.

\begin{figure}[htbp]
\centering
\includegraphics[width=0.52\textwidth]{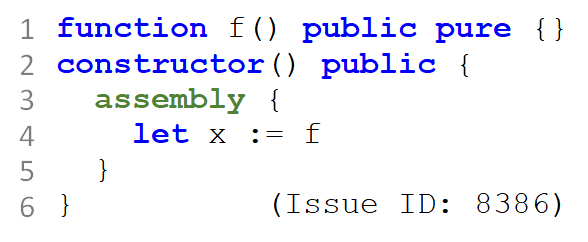}
\caption{An example for an incorrect analysis mechanism bug.}
\label{fig:Mechanism}
\end{figure}

Fig.~\ref{fig:Mechanism} represents a code example that triggers an incorrect analysis mechanism bug. The assembly block contains the statement \texttt{let x := f}, which references the function identifier \texttt{f} and assigns it to an assembly variable. However, inline assembly cannot directly access functions, which are high-level language constructs. Due to imperfections in the semantic analysis mechanism, the semantic analyzer erroneously allows this code to proceed to the subsequent code generation phase. The compiler ultimately crashes internally when encountering an unresolved function symbol during assembly node processing.

\begin{center}
\fcolorbox{black}{gray!25}{\parbox{0.97\linewidth}{
\textit{\noindent\textbf{Finding 6}: 
11.64\% of the typing-related bugs arise due to incorrect analysis mechanisms, particularly the inconsistency across the compiler's various analysis phases in interpreting the same language constructs and inadequate intermediate representations to fully capture the mapping relationship from syntax to semantics. 
}

\textit{
\noindent\textbf{Implication 6}: Compiler maintainers need to refine the architecture to ensure consistent interpretation of language constructs across all phases. Tool developers can build syntax-semantics consistency checkers to detect such inconsistencies early.
}}}
\end{center}

\subsection{Symbol and Scope Management Failures}
This category of bugs reflects deficiencies in the Solidity compiler's processes for identifier binding, scope management, and symbol table construction, preventing accurate association between symbols in code and their corresponding definitions. The core issue lies in the compiler's inconsistent handling of Solidity-specific scope hierarchies, including contract inheritance chains, library function bindings, and file import aliases.
When processing overloaded functions, the compiler may incorrectly prioritize parameter order over type matching precision during selection (\emph{e.g.}, \href{https://github.com/argotorg/solidity/issues/9752}{ID: 9752}), or lose discrimination accuracy when distinguishing between built-in attributes and user-defined functions (\emph{e.g.}, \href{https://github.com/argotorg/solidity/issues/2655}{ID: 2655}). Similarly, when handling symbols imported from external files, the mapping relationships between alias mechanisms and original symbols may break due to inconsistent symbol table states (\emph{e.g.}, \href{https://github.com/argotorg/solidity/issues/10956}{ID: 10956}).
A more fundamental problem is the compiler's lack of comprehensive contextual awareness. For instance, the parser frequently fails to correctly determine valid symbol scopes according to language specifications when encountering identifiers with identical names in nested scopes (\emph{e.g.}, \href{https://github.com/argotorg/solidity/issues/1731}{ID: 1731}), resolving function override and overload ambiguities within inheritance hierarchies (\emph{e.g.}, \href{https://github.com/argotorg/solidity/issues/9028}{ID: 9028}), 
or determining visibility of library functions bound through \texttt{using} statements (\emph{e.g.}, \href{https://github.com/argotorg/solidity/issues/13764}{ID: 13764}).
These resolution failures not only generate surface errors such as ``undefined symbol" 
but also trigger deeper issues including symbol table corruption and cascading type inference errors, ultimately revealing the compiler's weaknesses in managing symbol lifecycles and maintaining scope integrity.

\begin{figure}[htbp]
\centering
\includegraphics[width=0.92\textwidth]{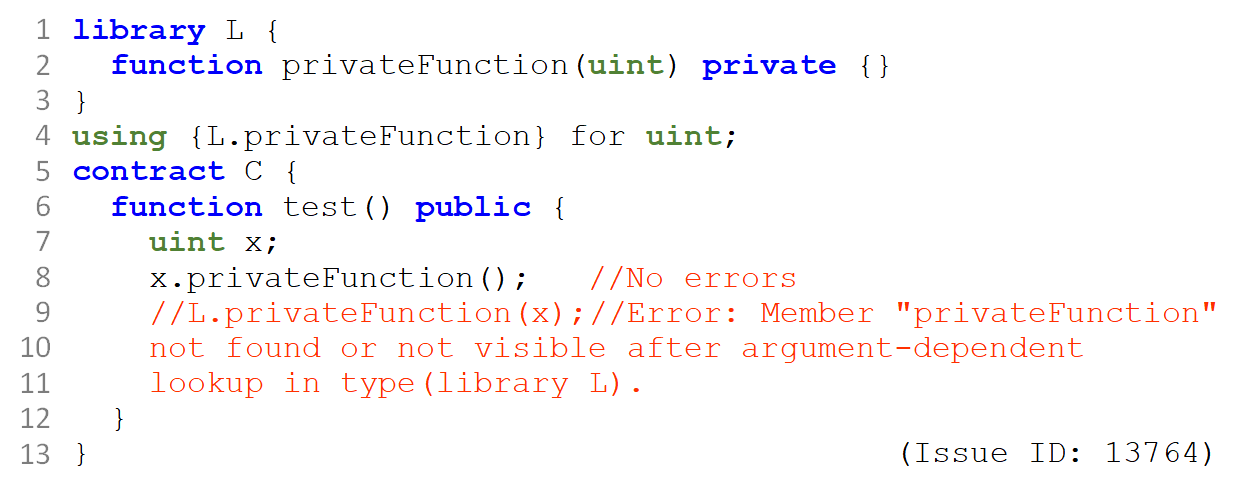}
\caption{An example for a symbol and scope management bug.}
\label{fig:resolution}
\end{figure}

Fig.~\ref{fig:resolution} gives a code example that represents a bug triggered by failures in symbol and scope management. Although \texttt{privateFunction} in library \texttt{L} is declared as private, the compiler incorrectly permits its binding to a type at the global scope through the \texttt{using} statement. This violates the scope rules governing private functions, resulting in the unintended exposure of a function that should remain restricted to internal library access. Consequently, contract \texttt{C} can invoke this private method via \texttt{x.privateFunction()}. However, when attempting direct invocation through the library name as \texttt{L.privateFunction(x)}, the compiler correctly denies access. This contradictory behavior reveals inconsistencies in the compiler's symbol visibility verification mechanism, particularly demonstrating scope management failures during the processing of \texttt{using} declarations.

\begin{center}
\fcolorbox{black}{gray!25}{\parbox{0.97\linewidth}{
\textit{\noindent\textbf{Finding 7}: 
The current Solidity compiler has limitations in managing symbol lifecycles and scopes when handling complex scope hierarchies, leading to 12.33\% of all bugs.
}

\textit{
\noindent\textbf{Implication 7}: Language designers should consider simplifying scope rules, and IDE developers need to implement more precise symbol resolution engines.
}}}
\end{center}

\subsection{AST and IR Transformation Errors}
Transformation errors during AST and IR processing reveal semantic equivalence violations in the Solidity compiler's code restructuring and intermediate representation generation. The essence of these bugs lies in the compiler's failure to maintain logical consistency when transforming, optimizing, or simplifying abstract syntax trees into lower-level representations (such as Yul IR). These issues manifest as fundamental disconnections between transformation algorithms and Solidity-specific semantic rules. 
For instance, during optimization phases (\emph{e.g.}, in the Yul optimizer), transformation logic for unchecked arithmetic operations produces behavioral discrepancies compared to traditional code generators, causing identical code to yield contradictory results across different compilation paths (\emph{e.g.}, \href{https://github.com/argotorg/solidity/issues/11631}{ID: 11631}).
More profound problems stem from broken context propagation between compiler phases. The transformation processes frequently lose critical semantic information (such as constructor visibility and contract type constraints), causing subsequent analysis stages (\emph{e.g.}, SMT model checking) to reason based on incorrect premises, which ultimately triggers type conversion exceptions or solver failures (\emph{e.g.}, \href{https://github.com/argotorg/solidity/issues/1364}{ID: 1364}).
These transformation errors expose integration defects between intermediate representation layers and semantic constraint systems.

\begin{figure}[htbp]
\centering
\includegraphics[width=0.62\textwidth]{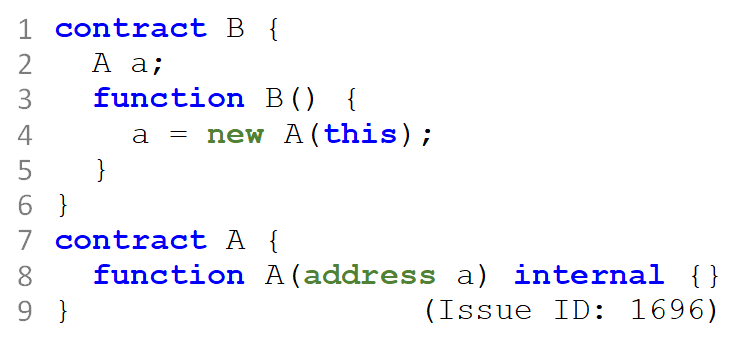}
\caption{An example that triggers an AST transformation bug.}
\label{fig:ast}
\end{figure}

Fig.~\ref{fig:ast} gives a code example that triggers an AST transformation bug. When the constructor of contract A is declared as \texttt{internal}, this critical information fails to be properly recorded and propagated in the AST. Consequently, during the subsequent code generation phase, when traversing the AST and encountering the \texttt{new A()} node, the compiler cannot determine through the AST that this creation operation is illegal. It thus erroneously attempts to generate bytecode for it, ultimately failing to find the corresponding compilation result and throwing an internal exception.

\begin{center}
\fcolorbox{black}{gray!25}{\parbox{0.97\linewidth}{
\textit{\noindent\textbf{Finding 8}: 
Type information loss during AST and IR transformations constitutes a fatal yet subtle problem, leading to 8.91\% of all bugs. 
}

\textit{
\noindent\textbf{Implication 8}: Compiler maintainers need to establish type-preserving transformation frameworks, and formal verification tools specifically targeting AST and IR transformations should be developed to validate transformation correctness.
}}}
\end{center}

\subsection{Defects in Error Handling Mechanisms}
This category of bugs reveals structural deficiencies in the compiler's error management mechanisms. Although the compiler can correctly detect violations in code, failures occur during the generation, transmission, and presentation of error information. The essence of this problem lies in the lack of robustness in the compiler's error handling pipeline, causing internal exceptions that should be converted into user-friendly error messages to instead escalate into unhandled Internal Compiler Errors (ICE) or misleading information.
When encountering complex error scenarios - such as overriding unimplemented virtual function (\emph{e.g.}, \href{https://github.com/argotorg/solidity/issues/5130}{ID: 5130}), illegal use of tuple assignments (\emph{e.g.}, \href{https://github.com/argotorg/solidity/issues/15075}{ID: 15075}), or event index parameters exceeding limits (\emph{e.g.}, \href{https://github.com/argotorg/solidity/issues/50}{ID: 50})
- the compiler fails to report errors accurately and instead crashes due to assertion failures or uncaught exceptions. Furthermore, the error message generation logic contains ambiguities that prevent it from providing precise guidance based on context. For instance, while \texttt{catch} clause parameters should mandatorily use the memory location, the compiler generically suggests choosing between storage or memory (\emph{e.g.}, \href{https://github.com/argotorg/solidity/issues/11405}{ID: 11405}).
These deficiencies in the error handling mechanism not only hinder developers from quickly locating problems but also expose the compiler's architectural neglect in managing exception paths.

\begin{figure}[htbp]
\centering
\includegraphics[width=0.62\textwidth]{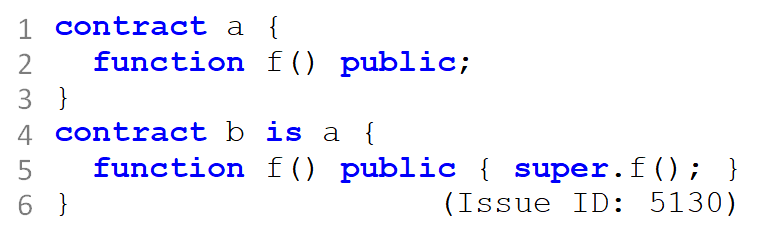}
\caption{An example that triggers an error handling mechanism bug.}
\label{fig:error}
\end{figure}

Fig.~\ref{fig:error} represents an example code that exposes a defect in the error handling mechanism. When contract \texttt{b} attempts to call the unimplemented virtual function from its parent contract \texttt{a} through \texttt{super.f()}, the compiler should provide clear compilation error messages to assist in fixing the code. However, it instead produces a confusing internal compiler error.

\section{Exposure Conditions for Typing-related Bugs (RQ3)}
\label{Exposure-Conditions}
This section presents our research results on the exposure conditions of typing-related bugs in the Solidity compiler. The exposure condition here refers to 
the language features from the perspective of the bug-triggering code, and each bug may simultaneously exhibit multiple exposure conditions. We first give categories of exposure conditions and then study the correlation between the root causes and exposure conditions.

\subsection{Categories of Exposure Conditions} 
 For the 146 typing-related bugs, the exposure conditions can be classified into seven categories as shown in Table~\ref{exposure}.

\begin{table}
\caption{Category of the exposure conditions for typing-related bugs.}
\begin{center}
\begin{tabular}{ l c c }
\toprule
\textbf{Exposure conditions categories} & \textbf{Instance} & \textbf{Percentage}  \\
\midrule
Type System Operations & 71 & 48.63\% \\
Data Location and Lifecycle & 51 & 34.93\% \\
Compilation Configuration & 34 & 23.29\% \\
Access Control & 30 & 20.55\% \\
External Interactions & 26 & 17.81\% \\
Special Data Structure Handling & 16 & 10.96\% \\
Using Inline Assembly & 7 & 4.79\% \\
\bottomrule
\end{tabular}
\label{exposure}
\end{center}
\end{table}

\subsubsection{Type System Operations}
These exposure conditions manifest primarily during the compiler's execution of specific operations including type definition, declaration, conversion, and composition. The problems concentrate in several critical dimensions.
In custom types and aliases handling, the compiler may fail to properly process visibility rules for User-Defined Value Types (UDVT), rendering their wrapper constructors inaccessible in specific contexts (\emph{e.g.}, \href{https://github.com/argotorg/solidity/issues/11955}{ID: 11955}). Alternatively, when applying \texttt{using} declarations to bind operators to custom types, internal errors may be triggered due to unresolved internal conversion logic (\emph{e.g.}, \href{https://github.com/argotorg/solidity/issues/15113}{ID: 15113}).
Declaration order dependencies similarly provoke issues. For example, when contracts reference internal constructors that haven't been fully resolved, the incomplete type information can cause compiler crashes (\emph{e.g.}, \href{https://github.com/argotorg/solidity/issues/1696}{ID: 1696}).
Regarding illegal type declarations, the compiler inadequately intercepts semantically invalid definitions. This includes permitting declarations of self-referencing constants (\emph{e.g.}, \href{https://github.com/argotorg/solidity/issues/1436}{ID: 1436}) or variables of library types (\emph{e.g.}, \href{https://github.com/argotorg/solidity/issues/7625}{ID: 7625}) --- errors that should be captured during compilation but instead directly induce internal exceptions.
Type conversion processes represent frequent failure points due to missing rules or implementation errors. Notable instances include the compiler mistakenly attempting to convert the special \texttt{super} keyword to a regular contract type (\emph{e.g.}, \href{https://github.com/argotorg/solidity/issues/9596}{ID: 9596}), or implicit conversions from string literals to fixed-size byte types causing SMT solver logic errors (\emph{e.g.}, \href{https://github.com/argotorg/solidity/issues/14792}{ID: 14792}).
The most complex scenarios occur in type composition. The compiler readily enters infinite loops when processing recursive structures such as structs containing self-referential mappings (\emph{e.g.}, \href{https://github.com/argotorg/solidity/issues/9443}{ID: 9443}). Moreover, for composite types containing mappings as function parameters or return values, the compiler often crashes during encoding size calculation due to conflicts in type system constraints (\emph{e.g.}, \href{https://github.com/argotorg/solidity/issues/4260}{ID: 4260}, \href{https://github.com/argotorg/solidity/issues/9957}{9957}).

\begin{center}
\fcolorbox{black}{gray!25}{\parbox{0.97\linewidth}{
\textit{\noindent\textbf{Finding 9}: 
Among the 71 bugs related to type system operations, type conversion (25 cases) and type composition (23 cases) account for the majority. The compiler's type system proves particularly susceptible to inference failures when Solidity code involves multi-level type nesting, mixed usage of custom and elementary types, or chains of implicit type conversions.
}

\textit{
\noindent\textbf{Implication 9}: 
{For compiler testing, it is recommended to develop type-aware fuzzing tools that automatically generate test cases containing complex structures such as nested types and custom types to comprehensively cover edge scenarios.
}
}}}
\end{center}

\subsubsection{Data Location and Lifecycle}
These exposure conditions focus on the rules governing data positioning, transfer, and lifecycle management across \texttt{storage}, \texttt{memory}, and \texttt{calldata}. The issues primarily manifest across multiple dimensions: memory management, referencing methods, composite type passing, and destructuring assignments.
In memory management, the compiler must accurately define data locations. However, bugs frequently cause rule conflicts, such as misinterpreting string literals as \texttt{storage} types instead of their default \texttt{memory} type (\emph{e.g.}, \href{https://github.com/argotorg/solidity/issues/4535}{ID: 4535}).
Improper handling of referencing methods can also trigger errors. For instance, compilation may fail when indirectly invoking member functions (\emph{e.g.}, \texttt{.pop}) on storage arrays (\emph{e.g.}, \href{https://github.com/argotorg/solidity/issues/10870}{ID: 10870}).
Regarding composite data type passing, the compiler proves particularly vulnerable when encoding and passing complex types like structs and arrays. A typical case involves infinite recursion when passing structs containing mappings as \texttt{memory} parameters (\emph{e.g.}, \href{https://github.com/argotorg/solidity/issues/9443}{ID: 9443}).
Bugs in destructuring assignment mechanisms predominantly emerge during tuple processing. When element counts mismatch between assignment operands (\emph{e.g.}, \href{https://github.com/argotorg/solidity/issues/4679}{ID: 4679}) or nested tuple structures are abnormal (\emph{e.g.}, \href{https://github.com/argotorg/solidity/issues/8450}{ID: 8450}), the compiler fails to robustly perform pattern matching, triggering internal errors.
Finally, while illegal assignment operations should be statically intercepted, the compiler may tacitly permit invalid operations. A representative example includes using compound assignment operators with tuples (\emph{e.g.}, \href{https://github.com/argotorg/solidity/issues/1720}{ID: 1720}).

\subsubsection{Compilation Configuration}
These exposure conditions originate from unexpected behaviors in specific stages of the compiler's internal pipeline when particular compilation options or experimental features are enabled. Such issues typically remain unexposed under default configurations.
When Yul optimization is enabled, the compiler employs a new intermediate representation for low-level optimizations. However, the optimization process may fail to correctly handle semantics of certain advanced Solidity constructs. For instance, when performing negation operations on the minimum value of signed integers, the Yul backend generates bytecode inconsistent with the conventional path, causing otherwise successful deployments to unexpectedly revert in optimized mode (\emph{e.g.}, \href{https://github.com/argotorg/solidity/issues/11631}{ID: 11631}). 
Similarly, when performing bit-shift operations on not fully supported fixed-point number types under Yul optimization, the compiler directly triggers internal errors instead of providing clear diagnostic messages (\emph{e.g.}, \href{https://github.com/argotorg/solidity/issues/10895}{ID: 10895}).
In scenarios with experimental ABIEncoderV2 enabled, the compiler gains capability to encode complex types, though its implementation remains unstable. For example, after activating this feature, processing nested structure type definitions can directly trigger compiler crashes (\emph{e.g.}, \href{https://github.com/argotorg/solidity/issues/5048}{ID: 5048}).
Additionally, when contracts with enabled and disabled ABIEncoderV2 attempt mutual calls, internal errors occur due to encoder version mismatches (\emph{e.g.}, \href{https://github.com/argotorg/solidity/issues/9270}{ID: 9270}).
The most complex scenarios emerge when enabling SMTChecker for formal verification. This tool requires transforming Solidity code into mathematical logic models, a process particularly susceptible to failures in edge cases. For instance, when mapping keys are of boolean type or require implicit conversion from string literals, the SMT solver throws logic error exceptions due to inability to handle specific type combinations (\emph{e.g.}, \href{https://github.com/argotorg/solidity/issues/14791}{ID: 14791}, \href{https://github.com/argotorg/solidity/issues/14792}{14792}). 
Similarly, when processing wrapping and unwrapping operations of User-Defined Value Types (UDVT), internal compiler errors occur as the SMT solver fails to track their underlying type conversion relationships (\emph{e.g.}, \href{https://github.com/argotorg/solidity/issues/15113}{ID: 15113}).

\begin{center}
\fcolorbox{black}{gray!25}{\parbox{0.97\linewidth}{
\textit{\noindent\textbf{Finding 10}: 
Among the 34 compilation configuration-related bugs, enabling SMTChecker (18 cases) and ABIEncoderV2 (13 cases) constitutes the primary risk. Furthermore, combining different compilation configuration options tends to produce more unpredictable bugs. 
}

\textit{
\noindent\textbf{Implication 10}: 
{A testing matrix incorporating features such as SMTChecker, ABIEncoderV2, and Yul optimization can be constructed to automatically execute cross-configuration combinatorial testing. Special attention should be paid to the interaction boundaries between experimental features and the traditional type system, thereby proactively exposing configuration-related risks.
}
}}}
\end{center}

\subsubsection{Access Control}
The exposure conditions for this category of bugs fundamentally arise when code triggers the compiler's verification process for function or data accessibility rules. These issues primarily manifest in two typical scenarios involving function visibility constraints and inheritance mechanisms.
In terms of function visibility, the compiler must strictly enforce access restrictions according to visibility modifiers (\texttt{public}, \texttt{private}, \texttt{internal}, \texttt{external}). However, bugs may cause these rules to be violated. For instance, the compiler might erroneously permit non-public state variables to use the \texttt{override} modifier in invalid attempts to ``override" parent contract functions (\emph{e.g.}, \href{https://github.com/argotorg/solidity/issues/9065}{ID: 9065}). 
Alternatively, the compiler may fail to correctly enforce the private visibility constraints of library functions, unintentionally allowing external contracts to call private library functions (\emph{e.g.}, \href{https://github.com/argotorg/solidity/issues/13764}{ID: 13764}).
Regarding inheritance constraints, the compiler must ensure logical consistency in type and function accessibility throughout inheritance hierarchies. Nevertheless, this safeguard may fail in complex scenarios. For example, the compiler might incorrectly treat external functions in libraries as inheritable members available for invocation (\emph{e.g.}, \href{https://github.com/argotorg/solidity/issues/2739}{ID: 2739}), or when a child contract fails to properly provide parameters required by a parent contract's constructor, the integrity of the inheritance chain may be compromised, triggering internal compiler errors (\emph{e.g.}, \href{https://github.com/argotorg/solidity/issues/4879}{ID: 4879}).

\subsubsection{External Interactions}
 These exposure conditions manifest during operations involving external interactions such as cross-contract calls, library functions, and higher-order functions. When processing these constructs, the compiler frequently exhibits deficiencies in semantic analysis and code generation. The fundamental challenge resides in accurately resolving and executing operations that extend beyond the current contract's execution context.
In scenarios involving library or API calls, the compiler must precisely translate calling conventions for both built-in functions and external libraries. However, critical functionality may fail under specific circumstances. For instance, when employing \texttt{abi.encode()} to encode a function reference with an associated gas limit, the compiler fails to properly process this compound expression, resulting in internal errors (\emph{e.g.}, \href{https://github.com/argotorg/solidity/issues/8590}{ID: 8590}). Furthermore, the compiler may incorrectly assess the state-modifying potential of external calls. A notable example occurs when correctly utilizing \texttt{staticcall} with explicit gas specifications within view functions, the compiler erroneously classifies the operation as potentially state-modifying, thereby violating the view semantics (\emph{e.g.}, \href{https://github.com/argotorg/solidity/issues/6901}{ID: 6901}).
At the higher-order function level, where functions are treated as parameters or return values, the compiler's type system encounters increased strain. For example, when passing an external function accepting \texttt{bytes} parameters as a callback, the compiler incorrectly determines \texttt{bytes} to be incompatible with \texttt{calldata} types (\emph{e.g.}, \href{https://github.com/argotorg/solidity/issues/1415}{ID: 1415}). 
Even sophisticated utilities like \texttt{abi.encodeCall} may fail when managing implicit data location conversions of function parameters (such as from \texttt{memory} to \texttt{calldata}) due to inflexible type-checking rules (\emph{e.g.}, \href{https://github.com/argotorg/solidity/issues/12718}{ID: 12718}).

\subsubsection{Special Data Structure Handling}
These exposure conditions originate from that the compiler's internal algorithms reach their capability boundaries during type inference, memory computation, or recursive processing when encountering unconventional, extreme-scale, and deeply nested data layouts. Such problems primarily concentrate in three typical scenarios.
First, when processing data with special numerical properties, the compiler's fundamental arithmetic logic may fail. Examples include performing negative exponent operations on signed integers (\emph{e.g.}, \href{https://github.com/argotorg/solidity/issues/9548}{ID: 9548}), or returning multi-dimensional fixed-size arrays with zero-length inner dimensions (\emph{e.g.}, \href{https://github.com/argotorg/solidity/issues/5054}{ID: 5054}).
Second, 
the compiler becomes highly susceptible to exceeding its internal integer representation limits (\emph{e.g.}, \texttt{u256}) when processing enormous numerical literals. Whether handling massive hexadecimal literals (\emph{e.g.}, \href{https://github.com/argotorg/solidity/issues/5052}{ID: 5052})
or variable declarations exceeding storage limits (\emph{e.g.}, \href{https://github.com/argotorg/solidity/issues/15899}{ID: 15899}), these situations cause compiler crashes during static computation phases due to numerical overflow or resource exhaustion.
Finally, for nested data structures, the compiler faces significant challenges in type resolution and memory management when processing multiply nested types. 
These include structs containing nested mappings and multi-dimensional dynamic arrays.
A typical case involves early ABI encoders directly crash due to insufficient support for nested structs (\emph{e.g.}, \href{https://github.com/argotorg/solidity/issues/5048}{ID: 5048}).

\subsubsection{Using Inline Assembly}
This exposure condition relates to the practice of embedding assembly code within program logic. It stems from the semantic disconnect between the compiler's type system and scope management and storage access mechanisms when processing embedded low-level assembly code alongside high-level Solidity semantics. When developers directly manipulate storage variables within assembly blocks, the compiler may fail to properly interpret their high-level intentions. For example, exceptions can occur when accessing the underlying layout of storage arrays through the \texttt{slot}  keyword (\emph{e.g.}, \href{https://github.com/argotorg/solidity/issues/9618}{ID: 9618}). 
Moreover, the compiler offers quite limited type support within inline assembly environments. Attempts to process Solidity types not fully supported in assembly (\emph{e.g.}, fixed-point numbers) lead to failures due to unsuccessful type category checks (\emph{e.g.}, \href{https://github.com/argotorg/solidity/issues/8412}{ID: 8412}). 
Scope and symbol resolution present additional weaknesses.
Assembly blocks cannot correctly identify constants defined after their declaration (\emph{e.g.}, \href{https://github.com/argotorg/solidity/issues/8881}{ID: 8881}), and referencing self-referential constants may even cause the compiler to enter infinite loops (\emph{e.g.}, \href{https://github.com/argotorg/solidity/issues/10732}{ID: 10732}).
These instances collectively reveal the compiler's fundamental challenges in isolating high-level contract logic from low-level assembly instructions.

\subsection{Correlation between the Root Causes and Exposure Conditions}
\label{Correlation between the Root Causes and Exposure Conditions}

To understand the relationship between the root cause of a typing-related bug and its exposure condition, following previous empirical studies about software bugs (\citealt{10.1145/1181309.1181314, 10.1145/3297858.3304069, performance}), we calculate a statistical metric named \emph{lift}. For a root cause category \emph{A} and a exposure condition category \emph{B}, the \emph{lift} of them is denoted by \emph{lift(AB)} and calculated as \emph{lift(AB)} = $\frac{P(AB)}{P(A)P(B)} $, where \emph{P(AB)} denotes the probability that a typing-related bug is caused by \emph{A} and exposed under condition \emph{B}. If the calculated value \emph{lift} is equal to 1, then the root cause \emph{A} is independent with the exposure condition \emph{B}. If the calculated \emph{lift} value is larger than 1, then the root cause \emph{A} and the exposure condition \emph{B} are positively correlated. The positive correlation means that if a typing-related bug is caused by \emph{A}, then it is more likely to be exposed under condition \emph{B}. If the calculated \emph{lift} value is smaller than 1, then \emph{A} and \emph{B} are negatively correlated.

Table ~\ref{liftRTEXP} presents the result, and we can get the following observations from it. First, the root cause category ``Core Type Operation Defects'' has positive correlations with the exposure condition categories ``Type System Operations'', ``Compilation Configuration'', and ``Special Data Structure Handling''. 
For its subcategories, the subcategory ``Incorrect Type Conversion'' has positive correlations with the exposure condition categories ``Type System Operations'' and ``Compilation Configuration'', 
the subcategory ``Incorrect Type Comparison and Type-related Constraint Calculation'' has a strong positive correlation with the exposure condition category ``Special Data Structure Handling'' (with a value of \texttt{3.04}),
and the subcategory ``Incorrect Type Inference'' has a strong positive correlation with the exposure condition category ``Using Inline Assembly'' (with a value of \texttt{3.48}). 
Second, the root cause category ``Semantic Analysis Defects'' has positive correlations with all exposure condition categories except ``Type System Operations'' and ``Special Data Structure Handling''. 
For its subcategories, the subcategory ``Missing Validation Checks'' has positive correlations with the exposure condition categories ``Data Location and Lifecycle'' and ``External Interactions'', 
and the subcategory ``Incorrect Analysis Mechanisms'' has a strong positive correlation with the exposure condition category ``Using Inline Assembly'' (with a value of \texttt{2.45}).
Third, the root cause category ``Symbol and Scope
Management Failures'' has positive correlations with the exposure condition categories ``Type System Operations'' and ``Access Control''.
Fourth, the root cause category ``AST and IR Transformation
Errors'' has a strong positive correlation with the exposure condition category ``Using Inline Assembly'' (with a value of \texttt{3.21}).
Finally, the root cause category ``Defects in Error Handling
Mechanisms'' has positive correlations with the exposure condition categories ``Data Location and Lifecycle'', ``Access Control'', and ``External Interactions''.
Overall, we can see that there exists a certain correlation between the root causes and the exposure conditions for typing-related compiler bugs. The correlation can guide testing efforts (\citealt{mttesting,mutationtesting,article,compileroptimization,6319229}) towards those that are more likely to expose bugs stemming from particular root causes, both in terms of prioritizing existing test cases and generating new ones.

\begin{table}
\caption{The statistical metric \emph{lift} between the root causes of typing-related bugs and their exposure conditions. The exposure conditions use the following shorthand notations: 
\textbf{TSO} (Type System Operations), \textbf{DLL} (Data Location and Lifecycle), \textbf{CC} (Compilation Configuration), \textbf{AC} (Access Control), \textbf{EI} (External Interactions), \textbf{SDSH} (Special Data Structure Handling), and \textbf{UIA} (Using Inline Assembly).}
\begin{center}
\begin{tabular}{@{} l c c c c c c c @{} c @{}}
\toprule
\textbf{} & \textbf{TSO} & \textbf{DLL} &\textbf{CC} &\textbf{AC} &\textbf{EI} &\textbf{SDSH} &\textbf{UIA}\\
\midrule
Core Type Operation Defects &1.19&0.85&1.05&0.94&0.69&1.60&0.73\\
\cline{1-1}
-\hspace{0.7em}Incorrect Type Conversion &1.68&0.95&1.43&0.90&0.62&0&0\\
\cline{1-1}
\makecell[l]{-\hspace{0.7em}Incorrect Type Comparison\\\hspace{1em}and Type-related Constraint\\\hspace{1em}Calculation} &0.75&0.76&0.72&0.97&0.75&3.04&1.39&\\
\cline{1-1}
-\hspace{0.7em}Incorrect Type Inference &1.03&1.19&1.07&0.81&0.94&0&	3.48\\
\hline
Semantic Analysis Defects &0.89&1.23&1.01&1.05&1.32&0.89&1.23\\
\cline{1-1}
-\hspace{0.7em}Missing Validation Checks &0.97&1.18&0.63&0.86&1.49&0.81&0.61\\
\cline{1-1}
\makecell[l]{-\hspace{0.7em}Incorrect Analysis\\\hspace{1em}Mechanisms} &0.73&1.35&1.77&1.43&0.99&1.07&2.45\\
\hline
\makecell[l]{Symbol and Scope\\Management Failures} &1.03&0.80&0.24&1.35&0.94&0&0\\
\hline
\makecell[l]{AST and IR Transformation\\Errors} &0.79&0.88&1.98&0.37&0.86&0.70&3.21\\
\hline
\makecell[l]{Defects in Error Handling\\Mechanisms} &0.59&1.23&0.61&1.39&1.60&0&0\\
\bottomrule
\end{tabular}
\label{liftRTEXP}
\end{center}
\end{table}

\section{Fix Strategies for Typing-related Bugs (RQ4)}
\label{fixstrategy}
This section presents our research results on the fix strategies for typing-related bugs in the Solidity compiler. We first give the categories of fix strategies and then investigate the correlation between the root causes and fix strategies.

\subsection{Categories of Fix Strategies}
The fix strategies for typing-related compiler bugs can be classified into four categories, as shown in Table~\ref{Fix}. Since certain bugs require more than one fix approach, they may belong to multiple categories simultaneously. 

\begin{table}
\caption{Category of the fix strategies for typing-related bugs.}
\begin{center}
\begin{tabular}{ l c c }
\toprule
\textbf{Fix strategies categories} & \textbf{Instance} & \textbf{Percentage}  \\
\midrule
Logic Correction & 95 & 65.07\% \\
Guard Check & 69 & 47.26\% \\
Error Handling Mechanism Modification & 53 & 36.30\% \\
Feature Addition & 9 & 6.16\% \\
\bottomrule
\end{tabular}
\label{Fix}
\end{center}
\end{table}


\subsubsection{Logic Correction}
This category of fixes directly addresses and rectifies core algorithms and internal logic within the compiler. The corrective measures include rewriting flawed algorithms, amending type inference rules, adjusting control flow, or updating code generation strategies, aiming to fundamentally resolve compilation issues such as incorrect bytecode output 
or internal state inconsistencies. Specific implementations involve correcting code generation logic that failed to clear higher-order bytes during type conversions (\emph{e.g.}, \href{https://github.com/argotorg/solidity/issues/446}{ID: 446}, \href{https://github.com/argotorg/solidity/issues/530}{530}), refining function overload resolution algorithms to accurately match parameter names and types (\emph{e.g.}, \href{https://github.com/argotorg/solidity/issues/9381}{ID: 9381}, \href{https://github.com/argotorg/solidity/issues/9752}{9752}), fixing offset calculations for slice operations and complex types (such as structs containing mappings) in memory computations (\emph{e.g.}, \href{https://github.com/argotorg/solidity/issues/9174}{ID: 9174}, \href{https://github.com/argotorg/solidity/issues/9443}{9443}), and implementing missing encoding logic for experimental features (\emph{e.g.}, ABIEncoderV2's support for nested structures, \href{https://github.com/argotorg/solidity/issues/2764}{ID: 2764}). These corrections ensure that the entire pipeline—from source code parsing to bytecode generation—operates in strict compliance with language specifications and developer expectations.

\subsubsection{Guard Check}
The essence of this category of fixes lies in proactively incorporating defensive validation steps along critical execution paths within the compiler. Its primary objective is not to alter core compilation logic or algorithms, but rather to establish an active safety net that performs mandatory screening of input data, internal states, or intermediate results for validity—prior to stages such as type system resolution and code generation. This effectively intercepts illegal inputs or exceptional conditions that would otherwise lead to internal compiler errors, stack overflows, or even crashes—such as infinite recursion caused by constant self-references (\href{https://github.com/argotorg/solidity/issues/10732}{ID: 10732}) 
and out-of-range numerical values (\href{https://github.com/argotorg/solidity/issues/8929}{ID: 8929})
— transforming these potential risks into manageable error-handling procedures before they propagate to vulnerable compiler components. Consequently, this significantly enhances compiler robustness, ensuring graceful degradation and meaningful feedback when encountering illegal code rather than producing unpredictable outcomes.
In addition, it enables the compiler to selectively compile code based on conditions, thereby generating shorter target programs, reducing memory footprint, and improving execution efficiency. 

\begin{center}
\fcolorbox{black}{gray!25}{\parbox{0.97\linewidth}{
\textit{\noindent\textbf{Finding 11}: 
Guard checks account for a substantial proportion (47.26\%) in bug fix strategies. 
}

\textit{
\noindent\textbf{Implication 11}: 
The prevalence of guard checks may mask deeper issues in type system design or compiler architecture, 
and compiler maintainers should treat guard checks as temporary measures while planning more fundamental architectural improvements.
}}}
\end{center}

\subsubsection{Error Handling Mechanism Modification}
This category of fixes aims to enhance the compiler's diagnostic capabilities, enabling earlier and more accurate detection and reporting of errors in user code. By modifying error handling mechanisms, various illegal patterns in source code—including but not limited to violations of typeing rules (\href{https://github.com/argotorg/solidity/issues/4670}{ID: 4670}) and structures non-compliant with language specifications (\href{https://github.com/argotorg/solidity/issues/3889}{ID: 3889})
-- are transformed from states that might originally trigger internal compiler errors, undefined behaviors, or silent acceptance into regular compilation errors or warnings that are accurately identified and clearly reported during early compilation stages. Specific measures include adding new compilation errors and warnings, or optimizing the precision and contextual relevance of existing error messages (such as explicitly identifying erroneous variables, locations, and expected types).
This prevents erroneous code from progressing to the code generation phase and safeguards the stability of the compilation process.

\begin{center}
\fcolorbox{black}{gray!25}{\parbox{0.97\linewidth}{
\textit{\noindent\textbf{Finding 12}: 
A sizable proportion (36.30\%) of fixes involve improvements to error handling mechanisms, indicating the compiler team's strong emphasis on developer experience. 
}

\textit{
\noindent\textbf{Implication 12}: 
Compiler documentation teams should develop best practice guidelines for type system usage based on common error patterns, and tool developers should create type-aware diagnostic engines that trace type inference processes and provide specific repair suggestions. 
}}}
\end{center}


\subsubsection{Feature Addition}
The core objective of this category of fixes is to extend the compiler's capabilities by introducing support for new language features or syntactic constructs. It goes beyond merely repairing bugs to actively enhance core components, such as the type system and code generation modules,
enabling the compiler to comprehend and process previously unsupported valid code. The modifications typically span multiple components, such as the type system, AST processing, and code generation modules.
A representative example includes implementing support for assigning string literals to \texttt{bytes32} constant declarations (\href{https://github.com/argotorg/solidity/issues/28}{ID: 28}).
Besides, this category encompasses critical language capability extensions such as implementing implicit conversions from \texttt{calldata} slices to \texttt{memory} arrays (\href{https://github.com/argotorg/solidity/issues/8991}{ID: 8991}), and permitting storage mappings to be passed as parameters to internal functions (\href{https://github.com/argotorg/solidity/issues/4635}{ID: 4635}).
Collectively, these fixes drive the evolution of the Solidity language, empowering developers with more powerful and flexible expression capabilities.

\vspace{2mm}
\noindent
\textbf{Additional Analysis of Fixes.}
To further understand the complexity of typing-related bugs, 
we conducted statistical analysis on the code scale and required time to fix these bugs.
First, we automatically measured the number of modified lines and affected source files for each fix. The results show that 80\% of bug fixes involve modifying 100 or fewer lines of code, while 6\% require fewer than 10 lines. The average modification spans 90 lines per fix, with a median of 42 lines. Additionally, most fixes affect only a limited number of files: 66\% of fixes involve 5 or fewer files, and 6\% modify just a single file.
Subsequently, we performed quantitative analysis of the time required to resolve these bugs. By extracting the creation and resolution dates for each bug, we calculated the fix duration as their difference. Statistics reveal that 63\% of bugs were fixed within one month, while 18\% required over three months for resolution. In terms of repair days, the median fix time is 17 days, with an average of 52 days.

\subsection{Correlation between the Root Causes and Fix Strategies}

To understand the relationship between the root cause of a typing-related bug and its fix strategy, we again calculate their \emph{lift} following the approach in Sec.~\ref{Correlation between the Root Causes and Exposure Conditions}. If the calculated value \emph{lift} is equal to 1, then the root cause \emph{A} is independent with the fix strategy \emph{B}. If the calculated \emph{lift} value is larger than 1, then the root cause \emph{A} and the fix strategy \emph{B} are positively correlated. The positive correlation means that if a typing-related bug is caused by \emph{A}, then it is more likely to be fixed by \emph{B}. If the calculated \emph{lift} value is smaller than 1, then \emph{A} and \emph{B} are negatively correlated.

Table ~\ref{liftRTFS} presents the result, and we can get the following observations from it.
First, the root cause category ``Core Type Operation Defects'' has no positive correlation with any fix strategies. 
For its subcategories, the subcategory ``Incorrect Type Conversion'' has a positive correlation with the fix strategy category ``Logic Correction'', 
the subcategory ``Incorrect Type Comparison and Type-related Constraint Calculation'' has positive correlations with the fix strategy categories ``Guard Check'' and ``Error Handling Mechanism Modification'', and the subcategory ``Incorrect Type Inference'' has a positive correlation with the fix strategy category ``Error Handling Mechanism Modification''.
Second, the root cause category ``Semantic Analysis Defects'' has positive correlations with all fix strategy categories except ``Logic Correction''.
For its subcategories, the subcategory ``Missing Validation Checks'' has positive correlations with the fix strategy categories ``Guard Check'' and ``Error Handling Mechanism Modification'',
and the subcategory ``Incorrect Analysis Mechanisms'' has a strong positive correlation with the fix strategy category ``Feature Addition'' (with a value of \texttt{3.82}).
Third, the root cause category ``Symbol and Scope
Management Failures '' has a strong positive correlation with the fix strategy category ``Feature Addition'' (with a value of \texttt{2.70}). 
Fourth, the root cause category ``AST and IR Transformation Errors'' has a positive correlation with the fix strategy category ``Logic Correction''. 
Finally, the root cause category ``Defects in Error Handling
Mechanisms'' has a positive correlation with the fix strategy category ``Logic Correction''.
Overall, we can see that there exists a high correlation between the root causes and the fix strategies for typing-related compiler bugs. The high correlation  indicates that it would be fruitful to investigate automated localization (\citealt{multiple-fault,guifault,yufse, flsurvey, yuguifl}) and fix (\citealt{6035728, yuEmSE, yutse,yang2025parameter,urli2018design,9393494,yangEMSE,xue2025}) techniques for typing-related bugs in the Solidity compiler.

\begin{table}
\caption{The statistical metric \emph{lift} between the root causes of typing-related bugs and their fix strategies. The fix strategies use the following shorthand notations: 
\textbf{LC} (Logic Correction), \textbf{GC} (Guard Check), \textbf{EHMM} (Error Handling Mechanism Modification), and \textbf{FA} (Feature Addition).}
\begin{center}
\begin{tabular}{ l c c c c c }
\toprule
\textbf{} & \textbf{LC} & \textbf{GC} &\textbf{EHMM} &\textbf{FA}\\
\midrule
Core Type Operation Defects &0.94&0.93&0.97&0.28\\
\cline{1-1}
-\hspace{0.7em}Incorrect Type Conversion &1.25&0.71&0.71&0.60\\
\cline{1-1}
\makecell[l]{-\hspace{0.7em}Incorrect Type Comparison and\\\hspace{1em}Type-related Constraint Calculation} &0.67&1.13&1.19&0\\
\cline{1-1}
-\hspace{0.7em}Incorrect Type Inference &0.90&0.71&1.15&0\\
\hline
Semantic Analysis Defects &0.93&1.20&1.24&1.59\\
\cline{1-1}
-\hspace{0.7em}Missing Validation Checks &0.81&1.43&1.46&0.48\\
\cline{1-1}
\makecell[l]{-\hspace{0.7em}Incorrect Analysis Mechanisms}  &1.18&0.75&0.81&3.82
\\
\hline
\makecell[l]{Symbol and Scope Management Failures}  &1.11&0.82&0.77&2.70\\
\hline
\makecell[l]{AST and IR Transformation Errors}  &1.18&0.81&0.64&0\\
\hline
\makecell[l]{Defects in Error Handling Mechanisms} &1.32&0.91&0.79&0\\
\bottomrule
\end{tabular}
\label{liftRTFS}
\end{center}
\end{table}

\section{Related Work}
\label{Related Work}
This section introduces research work closely related to this paper, including empirical study on compiler bugs and research on typing-related bugs.

\subsection{Empirical Study on Compiler Bugs}
Empirical studies on compiler bugs are crucial for improving compiler reliability (\citealt{xu2023silent, liu2023towards}). \cite{shen2021comprehensive} systematically analyze 603 bugs from three popular deep learning compilers, 
and identify 12 root causes and 6 manifestation ways. 
They propose guidelines for detecting and debugging deep learning compiler bugs.
\cite{romano2021empirical} conduct an empirical study on bugs in WebAssembly compilers, primarily analyzing 1, 054 bugs from Emscripten, AssemblyScript, and Rustc/Wasm-Bindgen. Through quantitative analysis of lifecycle, impact, and bug-triggering inputs alongside fix scale, they reveal challenges such as data type incompatibility, memory model discrepancies, and synchronous code transformation.
\cite{he2025bug} empirically study bugs in GCC and LLVM by analyzing 806 commits from 603 GitHub projects.
They extensively examine the impact scope and symptom manifestations
for compiler bugs, 
and summarize common fix strategies used by developers, such as modifying programs and restricting compiler versions.
\cite{xu2023silent} manually analyze 4,827 bug reports from GCC and Clang compilers, identifying and categorizing various compiler bugs. They detail how compiler optimizations inadvertently break program security properties, leading to severe vulnerabilities like information leaks and memory errors.
\cite{liu2023towards} conduct a large-scale empirical study on bugs in Python interpreters CPython and PyPy, 
and explore bug distributions, common symptoms, root causes, and bug fixes. 
By comparing bugs across interpreters, they propose recommendations for improving Python interpreter testing and debugging.
\cite{lu2024understanding} investigate 333 unique bugs in Java decompilers, 
and categorize these bugs, study their symptoms, and analyze their causes.
They design a differential testing framework for Java decompilers, providing strong support for future decompiler improvements.
\cite{zhong2025understanding} analyzes the impact of GCC and LLVM compiler bugs on practical development, offering a novel research perspective. By collecting 644 independent commits from real-world development, they study the effects of these bugs, including their impacts, how programmers circumvent them, and fix strategies.
\cite{ma2024towards} collect 533 bugs documented since the inception of the Solidity compiler. Through categorization of bug symptoms and causes, they provide an understanding of these bugs.

\subsection{Research on Typing-Related Bugs}
Research on typing-related issues in programming languages provides valuable insights for understanding type system complexities. \cite{chen2020empirical} introduce six dynamic typing-related practices in Python programs, which represent common but potentially risky uses of dynamic typing rules by developers. They conduct an empirical study on real-world Python systems, investigating whether these practices correlate with increased bug occurrence and how developers fix dynamic typing-related errors.
\cite{chen2024risky} develop a rule-based tool RUPOR that constructs an accurate type library to detect typing-related bugs. 
Using RUPOR, they empirically study 25 real-world projects and summarize common fix patterns, including inserting type checks to resolve typing-related errors.
\cite{chaliasos2022finding} propose an extensible program generator applicable to three popular JVM languages, 
and the generator generates programs that may trigger compiler type errors. 
They introduce two novel methods (type erasure mutation and type coverage mutation) that apply targeted transformations to input programs, revealing type inference and soundness compiler errors respectively.
\cite{troppmann2024typed} conduct a large-scale empirical study 
that systematically measures type annotation adoption rates and frequency of explicit type checks.
Through manual analysis, they further verify the feasibility of remote type confusion vulnerabilities caused by insufficient type checking in real-world projects.
\cite{xu2016python} propose a predictive analysis method for Python that detects bugs triggered by dynamic typing. 
They first collect execution traces, then model and reason about dynamic types and attribute sets of inputs by introducing symbolic variables, thereby encoding both executed traces and unexplored branches as symbolic constraints. By solving these constraints, ``neighboring" executions caused by input type variations can be systematically explored to discover type errors.
\cite{gong2015dlint} present DLint, a framework for detecting typing-related issues in JavaScript code through dynamic analysis.
Targeting errors caused by dynamic typing and implicit type conversion, they implement multiple checkers that capture type misuses difficult to detect via static analysis—such as string and undefined concatenation, and accessing undefined properties—during runtime.
\cite{pradel2015typedevil} develop TypeDevil, a dynamic analysis approach for JavaScript that warns developers about type inconsistencies. This approach collects type information through dynamic execution, constructing type graphs to identify runtime type inconsistencies in variables, properties, or functions. 
Applying TypeDevil to widely-used benchmarks and real-world web applications reveals 15 typing-related bugs.

\section{Conclusion}
This paper presents the first systematic empirical study of 146 typing-related bugs in the Solidity compiler
across 4 dimensions: symptoms, root causes, exposure conditions, and fix strategies. The 12 key findings obtained from the study show that typing-related bugs in the Solidity compiler are not only associated with the inherent weaknesses in core mechanisms such as type inference, conversion, and comparison, but are also deeply intertwined with blockchain-specific elements including storage models, gas optimization strategies, and formal verification components. We also discuss the implications of these findings, which shed light for developing targeted detection and prevention techniques for typing-related bugs.
Based on the findings and implications, we plan to develop specialized testing tools for the Solidity compiler's type system to enhance the reliability and security of smart contract compilation.




%
%

\bibliographystyle{spbasic}
\bibliography{references}

%
\end{document}